\documentclass[11pt,a4paper,fleqn]{article} \textheight=24.5 true cm \textwidth=17 true cm
\usepackage{epsfig}
\usepackage{graphicx}
\usepackage{amsmath}
\usepackage{cite}
\usepackage{hyperref}
\usepackage{rotating}
\usepackage{setspace}
\begin{document}
\begin{center}
{\Large\bf Electronic structure of helium atom in a quantum dot}\\
[8 mm]
Jayanta K. Saha$^{1}$, S. Bhattacharyya$^{2}$ and T. K. Mukherjee$^{3,*}$\\
[4mm]
$ ^1 $\emph{Indian Association for the Cultivation of Science, Jadavpur, Kolkata 700032, India}\\
$ ^2 $\emph{Acharyya Prafulla Chandra College, New Barrackpore, Kolkata 700131, India}\\
$ ^3 $\emph{Narula Institute of Technology, Agarpara, Kolkata 700109, India}\\[2mm]
$^{*}$ Corresponding author; E-mail : drtapanmukherjee@gmail.com
\end{center}
\begin{abstract}
Bound and resonance states of helium atom have been investigated inside a quantum dot by using explicitly correlated Hylleraas type basis set within the framework of stabilization method. To be specific, precise energy eigenvalues of bound $1sns$ ($^{1}S^{e}$) [$n=1-6$] states and the resonance parameters \textit{i.e.} positions and widths of $^{1}S^{e}$ states due to $2sns$ [$n=2-5$] and $2pnp$ [$n=2-5$] configuration of confined helium below $N=2$ ionization threshold of $He^{+}$ have been estimated. The two-parameter (Depth and Width) finite oscillator potential is used to represent the confining potential representing the quantum dot. It has been explicitly demonstrated that electronic structure properties become a sensitive function of the dot size. It is observed from the calculations of ionization potential that the stability of an impurity ion within quantum dot may be manipulated by varying the confinement parameters. A possibility of controlling the autoionization lifetime of doubly excited states of two-electron ions by tuning the width of the quantum cavity is also discussed here. \\\\
\textbf{PACS numbers:} 31.15.A$-$, 31.15.V$-$, 32.80.Zb, 37.10.Gh\\
\textbf{Key words:} Helium Atom, Variational Method, Quantum Dot, Correlation
\end{abstract}

\vspace{0.8cm}
\section{Introduction}
The subject of atomic systems under spatial confinement is of immense interest among the researchers since the advent of quantum mechanics as the spectral characteristics of atomic systems placed under different confinements change appreciably compared with those of free atoms \cite{jal, sab}. A number of phenomenological potentials have been used to model atoms within cavities \cite{cha}, atoms under pressure \cite{bhat}, impurities in quantum dots or nano crystals \cite{xx1}, nanopores \cite{xx2,xx3}, fullerenes \cite{lud} and foreign atoms in liquid helium environment \cite{liq} \textit{etc}. The study of quantum dots (QD) has got considerable attention in recent times due to its fundamental importance in theoretical researches as well as in fabricating new functional devices. The QD's (or artificial atoms), in general, contains several electrons subjected to an external confining potential and they show similar structural properties as compared to pure atoms. The structural changes of the impurity atoms inside QD's \textit{w.r.t.} the parameters of confining potentials provide huge physical insight about the interaction of the atom with its surroundings. Although the bound states of confined hydrogen and helium atoms have been studied extensively by several researchers \cite{jal, sab, bhat}, very few attempts have so far been made towards the quasi bound or resonance states of one electron impurity atom in an isolated QD \cite{xx4,xx5} and also in case of confined two electron systems \cite{bly, saj, gen}. Transformations of two-electron bound states to Feshbach and then to shape resonances depending upon a parameter of model rectangular well-type potential representing the QD have been studied by Bylicki \textit{et. al.} \cite{bly}. Sajeev \textit{et. al.} \cite{saj} and Genkin \textit{et. al.} \cite{gen} showed that the singly excited bound states of a two-electron atom become resonance states for appropriately chosen parameters of an external attractive spherical Gaussian type confining potential used to model the QD. \\
In the present work, we have considered a spherically symmetric finite oscillator potential \cite{cha,win,kim} of the type,
\begin{eqnarray}
V_{c}(r)=-V_{0}(1+c_{w}r)e^{-c_{w}r}
\end{eqnarray}                                                                                                                            for modeling the QD confinement. Here $V_{0}$ is the depth of the potential well and the cavity constant $c_{w}$ is defined as,
\begin{eqnarray}
c_{w}=\frac{1}{\Delta \sqrt{V_{0}}}
\end{eqnarray}                                                                                                                                                     where $\Delta$ is the width of the potential. By tuning the parameters $V_{0}$ and  $\Delta$ one can change the shape of the potential given by equation (1). Such type of two-parameter ($V_{0}$ and $\Delta$) potential provides much control and flexibility in modeling the size of a QD. When $r\rightarrow 0$ \textit{i.e.} near the center of QD , $V_{c}(r) \sim r^{2}$ and thus a harmonic nature is observed in the potential for a given cavity constant $c_{w}$. But for large `\textit{r}', it deviates from the harmonic behavior. In fact, The FO potential is quite similar in profile to that of Gaussian potential. At the same time, it facilitates the computation of matrix elements in a simple and efficient manner, especially when the Slater-type orbitals are used in constructing the wave function with appropriate boundary conditions for a confined system. This FO potential was used by Winkler \cite{win} to study the two-electron bound and resonant states of helium in QD where the electron correlation was not included initially in the optimized wave function. Even the inclusion of electron correlation could not remove the uncertainties in their calculations \cite{win}. Later, Kimani \textit{et. al.} \cite{kim} applied the restricted Hartree-Fock method to estimate the ground states of many-electron close-shell quantum dots modeled by the FO potential where the electron correlations were included approximately. Chakraborty and Ho \cite{cha} made a sophisticated approach to deal with this problem by expanding the wave function in single exponent Hylleraas type basis within the framework of stabilization method, but their work was restricted to only the lowest lying doubly excited resonance state $2s^{2}$ ($^{1}S^e$) of helium. It is worthwhile to mention that an appropriate knowledge of resonance structure of few-electron QD with and without a central impurity atom will help to understand the electron transport phenomena occurring in real semiconductor QDs \cite{bly}. \\
Under such circumstances, we have studied the resonance parameters of $^{1}S^{e}$ states originated from $2sns$ and $2pnp$ [$n=2-5$] configurations of QD confined helium below $N=2$ ionization threshold of $He^{+}$ in the framework of stabilization method \cite{man} by using explicitly correlated multi-exponent Hylleraas type basis set. This method was successfully employed by the present workers \cite{jks,jks1,jks2,jks3,jks4} for calculations of resonance parameters of different resonance states of the free and confined helium-like ions. In the present study, the resonance parameters of the states under consideration are estimated over a wide range of width ($\Delta$) for a fixed depth ($V_{0}$) of the FO potential. The energy values of bound $1sns$ ($^{1}S^{e}$) states [$n=1-6$] have also been been reported. Moreover, the positions of $ 1s $, $ 2s $ ($ ^2S $) and $ 2p $ ($ ^2P $) states of $He^+$ have been estimated for a comprehensive understanding about the structure of QD confined helium. The variation of ionization potential of QD confined \textit{He} with respect to the width of the FO potential have been studied. It has also been shown that the potential given by equation (1) breaks the orbital angular momentum ($l$) degeneracy in Coulomb field for the energy levels of hydrogen-like atoms. Finally, we have shown that for a fixed cavity depth ($V_{0}$), the widths of the resonance states show oscillatory behavior with respect to the width ($\Delta$) of the quantum cavity. It has been noted that for higher excited states, such oscillations are more pronounced. The paper is arranged as follows: A brief discussion on the present methodology is given in Section II, followed by a discussion on the results in Section III, and finally concluding in Section IV with a view towards further use of the present techniques in related studies of spatially confined atomic systems \textit{e.g.} QD, presuure confinement, SCP confinement \textit{etc}.
\section{Method}
For any $^{1}S$ state of even parity arising from two electrons having same azimuthal quantum number, the variational equation \cite{tkm} can be written as,
\begin{eqnarray}
\delta \int \left[ \left(\frac{\partial f}{\partial r_{1}} \right)^{2} + \left(\frac{\partial f}{\partial r_{2}}\right)^{2} + \left(\frac{1}{r_{1}^{2}} + \frac{1}{r_{2}^{2}}\right)\left(\frac{\partial f}{\partial \theta_{12}}\right)^{2} + 2 \left( V_{eff}-E \right)f^{2} \right] dV_{r_{1},r_{2},\theta_{12}} = 0
\end{eqnarray}
subject to the normalization condition,
\begin{eqnarray}
\int f^{2}dV_{r_{1},r_{2},\theta_{12}}=1
\end{eqnarray}
where the symbols used in equation (3) and equation (4) are the same as in reference \cite{tkm}. The effective potential is given by,
\begin{eqnarray}
V_{eff}=\sum_{i=1}^2 \left[-\frac{2}{r_{i}}+V_{c}\left(r_{i}\right)\right]+\frac{1}{r_{12}} 
\end{eqnarray}
The multi-exponent correlated wavefunction \cite{jks2} considered in the present calculation is expressed as,
\begin{eqnarray}
f(r_{1},r_{2},r_{12})&=&\sum_{i=1}^{9}\eta_{i}(1)\eta_{j}(2)\left[\sum_{l\geq 0}\sum_{m\geq 0}\sum_{n \geq 0}C_{lmn}r_{1}^{l}r_{2}^{m}r_{12}^{n}+ exchange \right] {\nonumber}\\
&+& \sum_{i}\sum_{j}\left[\eta_{i}(1)\eta_{j}(2)\sum_{l\geq 0}\sum_{m \geq 0}\sum_{n\geq 0}C_{lmn}r_{1}^{l}r_{2}^{m}r_{12}^{n}+ exchange \right]
\end{eqnarray}
where,
\begin{eqnarray}
\eta_{j}(i)=e^{-\sigma_{i}r_{j}}
\end{eqnarray}                                                                                                                                                 where, $\sigma$'s are the non-linear parameters. Here, $r_{1}$ and $r_{2}$ are the radial co-ordinates of the electrons and $r_{12}$ is the relative distance between them. In a multiexponent basis set, if there are $p$ number of non-linear parameters, then the number of terms in the radially correlated basis is $\frac{p(p+1)}{2}$ and, therefore, the dimension of the full basis ($N$) including angular correlation will be $\left[\frac{p(p+1)}{2}\times q \right]$, where $q$ is the number of terms involving $r_{12}$ \cite{jkb}. For example, as we have used here nine non-linear parameters, the number of terms in the radially correlated basis is 45 and with 10 terms involving different powers of $r_{12}$, the dimension of the full basis ($N$) becomes 450. 
The values of the non-linear parameters are taken in a geometrical sequence: $\sigma_{i}=\sigma_{i-1}\gamma$, $\gamma$ being the geometrical ratio \cite{mb}. The wavefunction can be squeezed or can be made more diffuse by changing the geometrical ratio ($\gamma$) keeping $\sigma_{1}$ constant throughout. To have a preliminary guess about the initial and final values of nonlinear parameter $\sigma$, we optimize the energy eigenvalues of $^{1}S^{e}$ states below $N = 1$ ionization threshold of $ He^+ $ by using Nelder-Mead procedure \cite{nm}. The energy eigenroots are then obtained by solving the generalized eigenvalue equation,
\begin{eqnarray}
\underline{\underline{H}}~\underline{C}=E\underline{\underline{S}}~\underline{C}
\end{eqnarray}                                                                                                                                                    
where, $\underline{\underline{H}}$ is the Hamiltonian matrix, $\underline{\underline{S}}$ is the overlap matrix and $E$'s are the energy eigenroots. The wavefunction is normalized for each width ($\Delta$) of the FO potential to account for the modified charge distribution inside the QD. Each energy eigenroot plotted against the geometrical ratio ($\gamma$) produces the stabilization diagram. Subsequently, we can calculate the density of resonance states from the inverse of tangent at different points near the stabilization plateau in the neighborhood of avoided crossings for each energy eigenroot. The plot of calculated density of resonance states versus energy for each eigenroot is then fitted to a standard Lorentzian profile. The best fit, \textit{i.e.}, with the least chi square ($ \chi^2 $) and the square of correlation ($ R^2 $) near unity yields the desired position ($ E_r $) and width ($ \Gamma $) of the resonance state.\\
For each width ($\Delta$) of the confining potential, the energy eigenvalues of $^2S$ and $^2P$ states of the confined one-electron $He^{+}$ ion is obtained by using Ritz variational technique considering the wavefunction as,
\begin{eqnarray}
\psi = \sum_{l} C_{l}r^{l}e^{-\eta r}
\end{eqnarray}                                                                                                                                  
Where, $\eta$'s are the nonlinear parameters and $C$'s are the linear variational coefficients. For $He^{+}$($ns$) states [$ n=1-2 $], we have considered 14-parameter basis set whereas for $He^{+}$($2p$) state we have taken 13 parameters in the basis. In both the cases, $l$ is ranging from 0 to 4. All calculations are carried out in quadruple precision. Atomic units have been used throughout unless otherwise specified.
\section{Results and Discussions} 
To construct the stabilization diagram corresponding to each width ($\Delta$) of the FO potential, repeated diagonalization of the Hamiltonian matrix in the Hylleraas basis set of 450 parameters is performed in the present work for 400 different values of $\gamma$ ranging from 0.63 $a.u.$ to 0.77 $a.u.$ For a depth $V_{0}$ = 0.2 $a.u.$ and a width $\Delta$ = 4.0 $a.u.$ of the confining potential, a portion of the stabilization diagram for $^{1}S^{e}$ states of confined helium below $N=2$ ionization threshold of $He^{+}$ is given in figure-1. It is evident from figure 1 that there exist two classes of states:
\begin{enumerate}
\item First few energy eigen-roots lying below $He^{+}(1s)$ [$ -2.184879 $ a.u.] level are insensitive with the variation of $\gamma$. This feature clearly suggests that these energy eigen-roots originating from $1sns$ configurations of QD confined helium are bound \textit{i.e.} stable against auto-ionization.
\item Energy eigen-roots lying between $He^{+}(1s)$ and $He^{+}(2s)$ [$ -0.607849$ a.u.] are sensitive with the variation in $\gamma$ and gives rise to flat plateau in the vicinity of avoided crossings of the energy eigenroots in the neighborhood of some particular energy values. This is a clear signature for the presence of $^{1}S^{e}$ resonance states of QD confined helium.
\end{enumerate}
The present calculated bound state energy eigenvalues ($-E$) of $1sns$ ($^{1}S^{e}$) [$n = 1-6$] states of $He$ as well as the $He^{+}(1s)$ energies for different cavity widths ($\Delta$) starting from a very low value of 0.001 a.u. (corresponds to almost a free case) to a high value of 1000.0 a.u. are illustrated in Figure 2. It is to be noted that for very small cavity width $\Delta =0.001$, the $1sns$ [$n=1-6$] energy eigenvalues of helium and the $He^{+}(1s)$ threshold energy are nearly identical to the corresponding energy eigenvalues of the free ions and they remain almost unaltered  upto the cavity width $\Delta =0.1$ a.u. We can see from equation (1) that, for $ \Delta \rightarrow 0 $, $ c_w \rightarrow \infty$ and thus, $ V_c \rightarrow 0 $ which produce no effect of confinement. In between $\Delta =0.1$ a.u. and 10.0 a.u., the energy eigenvalues of helium decrease monotonically and ultimately saturates  at ($E_{1sns}+2\times V_{0}$) a.u. In a similar fashion, the threshold energy $He^{+}(1s)$ saturates at ($E_{1s}+V_{0}$) a.u. This feature is physically consistent as we can note from equation (1) that for $\Delta \rightarrow \infty$, the cavity constant $c_{w}\rightarrow 0$, so that $V_{c}(r)\rightarrow -V_{0}$. Thus the one and two electron energy levels will undergo a downward shift by $V_{0}$ and 2$V_{0}$ respectively for $\Delta\rightarrow\infty$. The variation of the ionization potential (in eV) \textit{i.e.} the energy required to ionize one electron from the ground state ($1s^{2}$) of helium atom is plotted against the width ($\Delta$) of the cavity in Figure 3. In accordance with the variation of energy eigenvalues of helium and its one-electron subsystem, it is evident from figure 3 that, the IP is identical with the vacuum IP for low values of $\Delta$ while for high values of $\Delta$, it increases by an amount $V_{0}\sim 5.44$ eV. It is thus evident from figures 2 and 3 that rate of variation of energy values of the ions are significant when the size of the confining cavity is of the order of atomic dimensions. It is also remarkable that the stability of an impurity atom can be controlled by suitably tuning the size of a QD \textit{i.e.} the depth and width of the representing cavity.\\
An enlarged view of the stabilization diagram (given in figure 1) for $^{1}S^{e}$ states of $He$ within the energy range $-0.8$ a.u. to $-0.64$ a.u. is given in figure 4. The $^{1}S^{e}$ states of $He$ below $N=2$ ionization threshold of $He^{+}(2s)$ can arise due to $2sns$ and $2pn'p$ ($n, n' \geq 2$) configurations. From a closer look at figure 4, we can see that for a short range of $\gamma$, each eigenroot between $N = 1$ and $N = 2$ ionization thresholds of $He^{+}$ becomes almost flat in the vicinity of avoided crossings in the neighborhood of different energies. In order to calculate the exact resonance parameters, the density of states (DOS) $\rho (E)$ is calculated by evaluating the inverse of the slope at a number of points near these flat plateaus of each energy eigenroot using the formula \cite{jks} given by:
\begin{eqnarray}
\rho_{n}(E)=\left\vert\frac{\gamma_{i+1}-\gamma_{i-1}}{E_{n}(\gamma_{i+1}) -E_{n}(\gamma_{i-1})}\right\vert_{E_{n}(\gamma_{i})=E_{i}}
\end{eqnarray} 
The estimated DOS $\rho_{n}(E)$ is then fitted to the following Lorentzian form \cite{jks}
\begin{eqnarray}
\rho_{n}(E)=y_{0}+\frac{A}{\pi}\frac{\frac{\Gamma}{2}}{\left(E-E_{r}\right)^{2}+\left(\frac{\Gamma}{2}\right)^{2}}
\end{eqnarray}
where, $y_{0}$ is the baseline background, $A$ is the total area under the curve from the baseline, $E_{r}$ gives the position of the center of the peak of the curve and represents the full width of the peak of the curve at half maxima. Among different fitting curves for each eigenroot corresponding to a particular resonance state, the best fitting curve \textit{i.e.} with least $\chi^{2}$ and the square of correlation ($ R^2 $) closer to unity \cite{jks} leads to the desired resonance energy ($E_{r}$) and width ($\Gamma$). The evaluation of DOS following the fitting procedure has been repeated for each width of the confining potential ($\Delta$). For example, the calculated DOS and the corresponding fitted Lorentzian for the $2s^{2}$ ($^{1}S^{e}$) resonance state of $He$ below $He^{+}(1s)$ threshold for cavity width $\Delta=4.0$ a.u. (given in figure 5) yields resonance position $E_{r}$ at $-0.98163$ a.u. and width $ \Gamma $ = $6.9961 \times 10^{-3}$ a.u.\\
The estimated resonance energies of doubly excited $2sns$ [$n=2-5$] ($^{1}S^{e}$) and $2pnp$ [$n=2-5$] ($^{1}S^{e}$) states of helium and corresponding $2s$ and $2p$ threshold energies for the cavity depth $V_{0}$ = 0.2 a.u. and cavity width ($\Delta$) ranging from 0.001 a.u. to 1000 a.u. are  given in figure-6, while the variations of resonance energies ($E_{r}$) of $2pnp$ [$n=2-5$] ($^{1}S^{e}$) states and corresponding $2s$ and $2p$ threshold energies versus $\Delta$ are given in figure-7. We have noted the following points.
\begin{enumerate}
\item It is clear from figures-6 and 7 that for $\Delta$ = 0.001 a.u., the $He^{+}$ ($2s$) and $He^{+}$ ($2p$) states are degenerate and coincide with the energy value of $N=2$ ionization threshold of free $He^{+}$ ion. As $\Delta$ increases, the $He^{+}$ ($2s$) and $He^{+}$ ($2p$) states become non-degenerate. Initially, the $2s$ level of $He^+$ lies energetically below the $2p$ level for $ \Delta $ up to 0.5 a.u. At $ \Delta $ = 1.0 a.u., the $2s$ state moves above the $2p$ level. These results exhibits that an `\textit{incidental degeneracy}' takes place for $2s$ and $2p$ states of $He^+$ at some value of $ \Delta $ between 0.5 and 1.0 a.u. and then a `\textit{level crossing}' occurs between two states having different symmetry properties. Finally these states become degenerate again for $\Delta \geq 100.0 $ a.u. The incidental degeneracy for $He^{+}(2s)$ and $He^{+}(2p)$ states occur for $\Delta$ in the range $0.5\leq \Delta \leq 1.0$. Such incidental degeneracy and subsequent level crossing phenomenon are being noted earlier by Sen \textit{et. al.} \cite{sen} in case of cage confined hydrogen atom and by Bhattacharyya \textit{et al.} \cite{scp} in case of helium-like ions within strongly coupled plasma environment.  
\item It is seen from both figures-6 and 7 that all the resonance energies ($E_{r}$) are almost unaltered up to $\Delta$ = 0.5 a.u., then decrease rapidly up to $\Delta$ = 20.0 a.u. and ultimately saturates. For low values of $\Delta$ (say 0.001 a.u.) the resonance energies are identical with those of the free \textit{He} atom whereas for $\Delta$ = 1000.0 a.u. the resonance energies are equal to those of free \textit{He} atom plus 0.4 a.u. ($ i.e. $ 2.0$\times V_0$). Thus, for a given depth ($V_{0}$) of the finite oscillator potential, the variations of energies of the bound states and the resonance states of helium \textit{w.r.t.} the width of the cavity ($\Delta$) are nearly identical.
\end{enumerate}
The variation of widths ($\Gamma$) of $2sns$ and $2pnp$ ($^{1}S^{e}$) $[n=2-5]$ resonance states \textit{w.r.t.} $\Delta$ are given in figures-8 and 9 respectively. A closer look at the figures 8 and 9 leads us to the following observations.
\begin{enumerate}
\item In general, it can be argued that the variation of widths shows an oscillatory behavior which are more pronounced for the higher excited states. It is worthwhile to mention here that recently Chakraborty and Ho \cite{cha} also reported such oscillation of resonance width ($\Gamma$) for $2s^{2}$ ($^{1}S$) state of QD confined helium atom. This feature clearly indicates a possibility of controlling the autoionization lifetime of doubly excited states of two-electron ions by tuning the parameters of the confining FO potential representing the quantum dot.
\item The variations of widths of $2s^{2}$ and $2p^{2}$ ($^{1}S^e$) states with respect to $\Delta$ are exactly opposite in nature. For $^{1}S^e$ state originating from $2s^{2}$ configuration, the autoionization width first decreases in the range $0.1\leq\Delta\leq 1.0$ and after reaching the minima, it shows a large bump around $\Delta\simeq$ 6.0 a.u. After that it starts to decrease and finally the autoionization width saturates where it becomes equal to that of a free \textit{He} atom. In contrast, for $2p^{2}$ state, the autoionization width first increases for $0.1\leq\Delta\leq 1$  and then shows a large dip approximately at the same value of $\Delta$ for which the $2s^{2}$ state shows the bump.
\item The values of $\Delta$ corresponding to the largest bump in the values of autoionization widths ($\Gamma$) of $2sns$ states and the lowest dip for $2pnp$ states shift towards the higher values of the cavity width ($\Delta$) for higher excited states.  
\end{enumerate}
Inside the QD \textit{i.e.} due to the presence of the surrounding FO potential, the charge distribution of the impurity ion gets reoriented which produces the behavioral changes as compared to a free ion. The nodes or antinodes of the resonance wavefunction lie at the boundary of the QD cavity and the interference caused inside the cavity gives rise to the oscillatory behavior of the resonance widths \cite{cha, rev}. The number of nodes or antinodes of the wavefunction increases for high-lying resonance states and the oscillation becomes more prominent. 

\section{Conclusion}
Structural properties of He atom confined in a QD, efficiently modeled by a two-parameter weakly confining FO type potential, has been investigated in the framework of Stabilization method using explicitly correlated Hylleraas-type basis sets. It has been observed that the structure of the impurity ion is a sensitive function of the dot size. For very small values of the cavity width, the system behaves almost like a free ion whereas, for very high cavity widths, a constant shift equal to the depth of the potential are observed in the bound as well as resonance energies. When the dot size becomes comparable to the dimensions of the impurity atom, the effects are more pronounced and many remarkable behaviors such as increase in ionization potential, oscillations in the widths of two-electron resonance states, incidental degeneracy and subsequent level-crossing phenomena for one-electron ions are observed. The present work is expected to lead to future investigations on the autoionizing states of different angular momenta for QD confined two-electron systems.\\\\\\
\textbf{Acknowledgment}\\\\
TKM gratefully acknowledges financial support under grant number 37(3)/14/27/2014-BRNS from the Department of Atomic Energy, BRNS, Government of India. SB acknowledges financial support under Grant No. PSW-160/14-15(ERO) from University Grants Commission, Government of India.

\newpage
\begin{figure}
\includegraphics[scale=0.5]{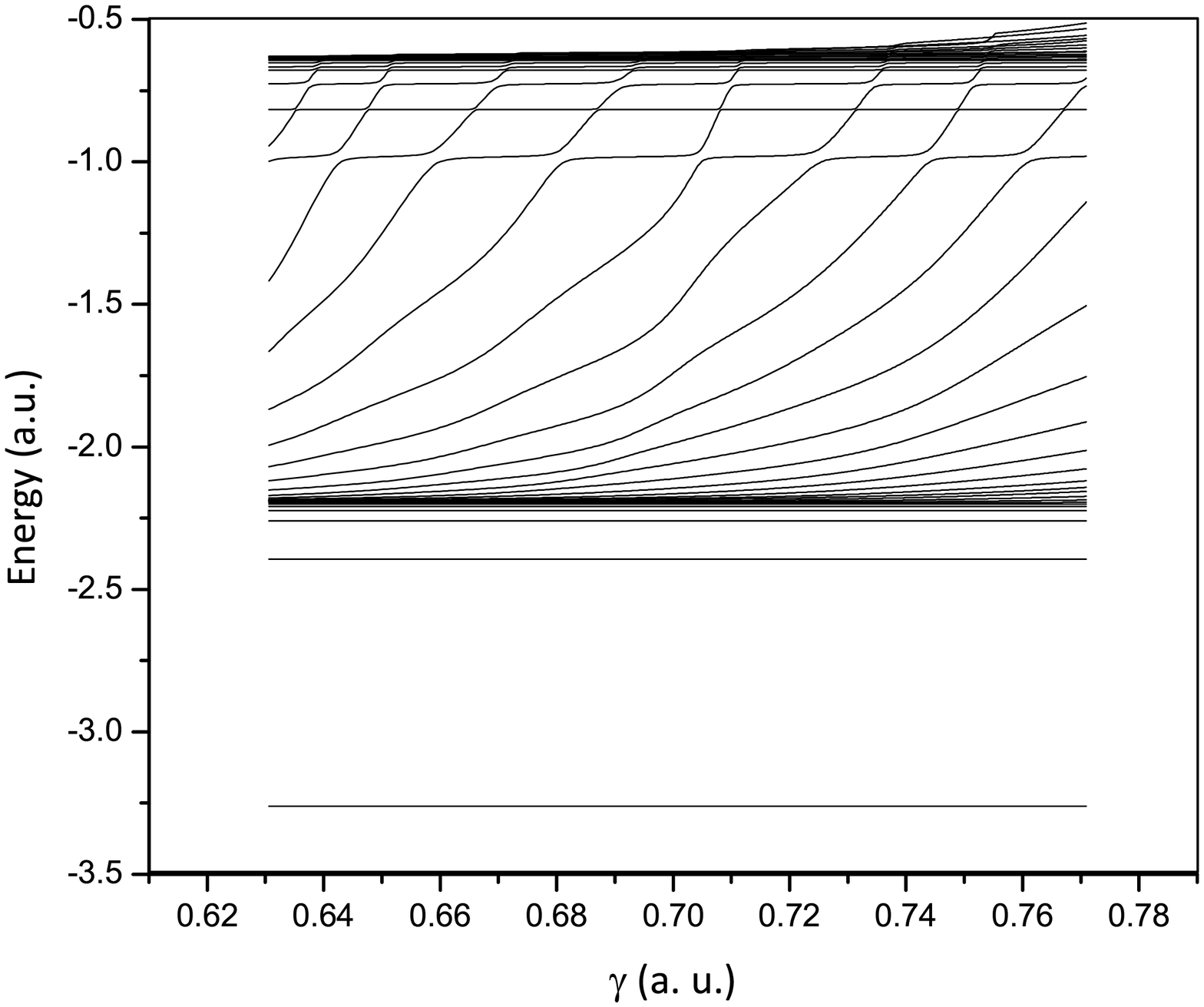}
\vspace{-0.5cm}
\caption{Stabilization diagram for $^{1}S^{e}$ states of helium atom under quantum cavity. Width of the cavity is set at 4.0 $a.u.$}
\end{figure}
\begin{figure}
\includegraphics[scale=0.5]{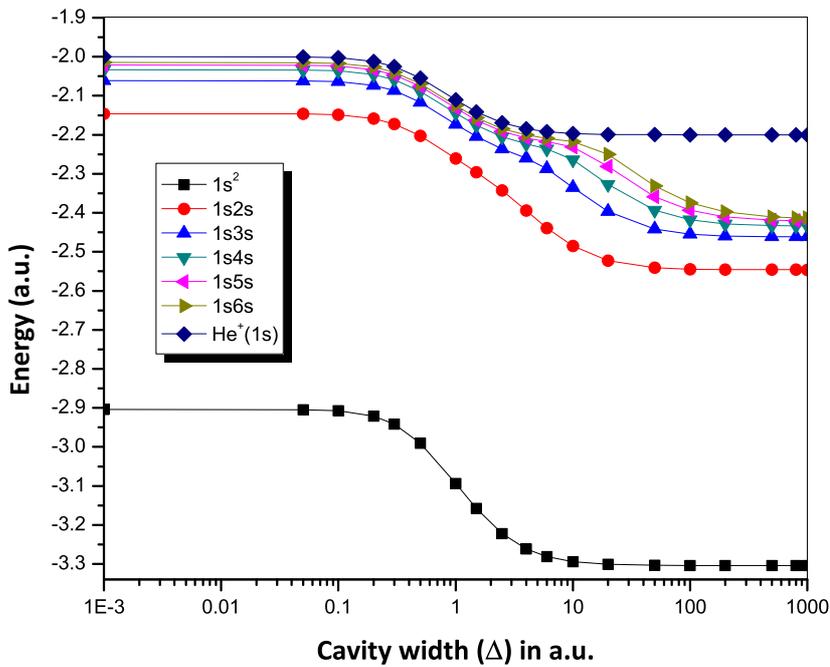}
\vspace{-0.5cm}
\caption{The variation of bound state energy eigenvalues w.r.t. the width of the cavity}
\end{figure}
\begin{figure}
\includegraphics[scale=0.5]{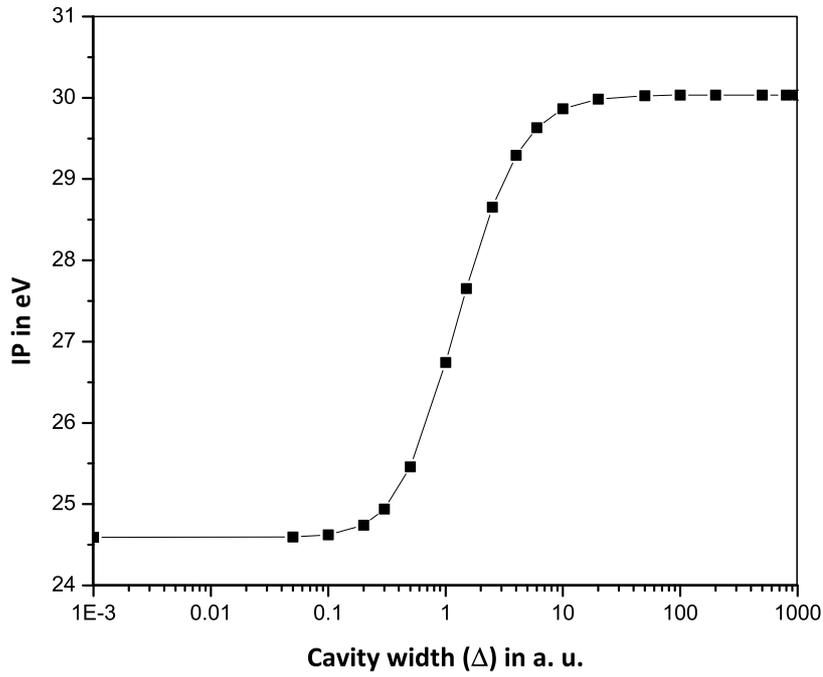}
\vspace{-0.5cm}\caption{The variation of IP w.r.t. the width of the cavity}
\end{figure}
\begin{figure}
\includegraphics[scale=0.5]{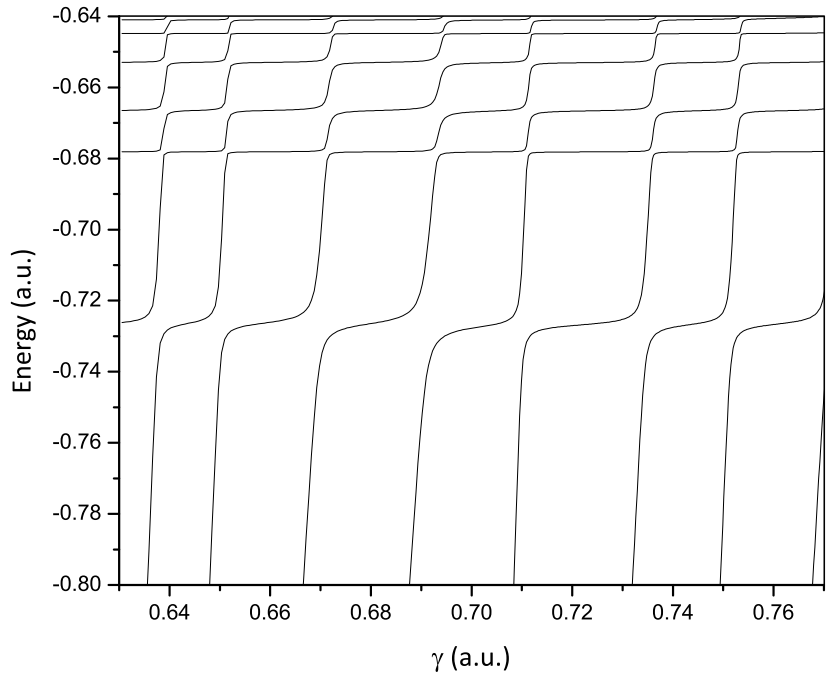}
\vspace{-0.5cm}
\caption{Enlarged view of the stabilization diagram for $^{1}S^{e}$ states of helium atom under quantum cavity in the energy range between -8.0 a.u. to -0.64 a.u. Width of the cavity is set at 4.0 $a.u.$}
\end{figure}
\begin{figure}
\includegraphics[scale=0.5]{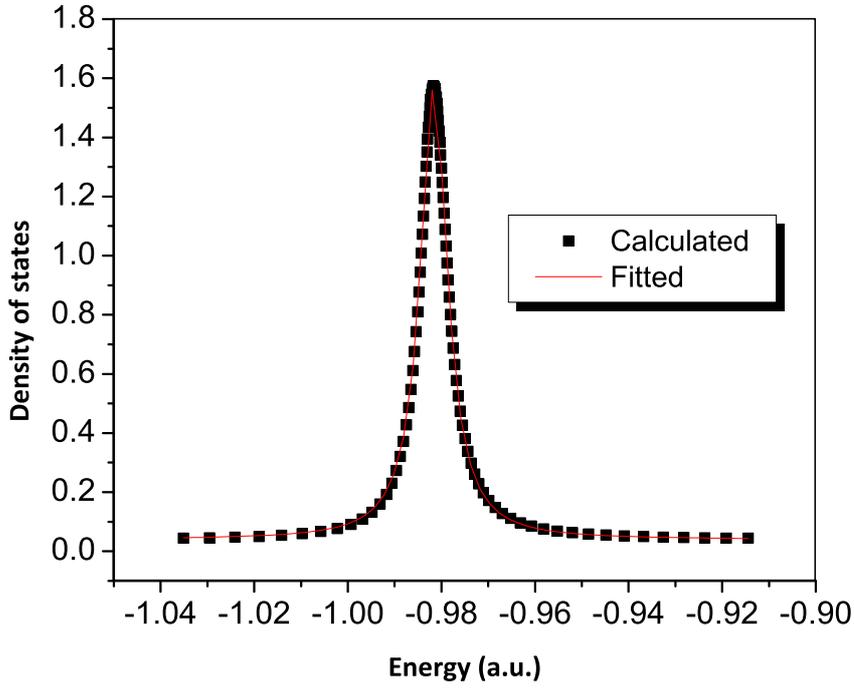}
\vspace{-0.5cm}
\caption{Density of states and fitted lorentzian for cavity width 4.0 a.u.}
\end{figure}
\begin{figure}
\includegraphics[scale=0.5]{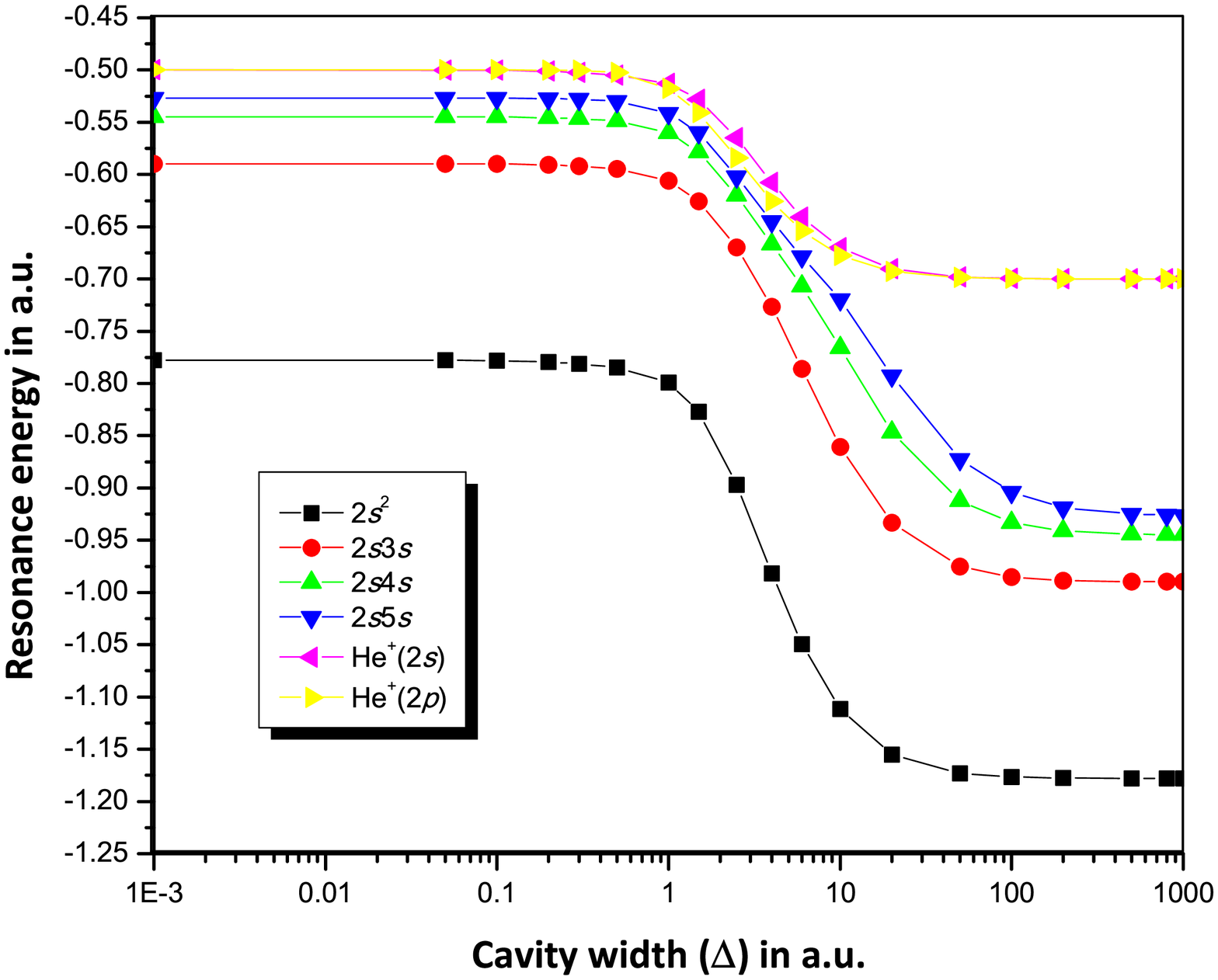}
\vspace{-0.5cm}
\caption{The variation of resonance energies ($E_{r}$) of $2sns$ [$n=2-5$] ($^{1}S^{e}$) states and corresponding $2s$ and $2p$ threshold energies with the cavity width ($\Delta$).}
\end{figure}
\begin{figure}
\includegraphics[scale=0.5]{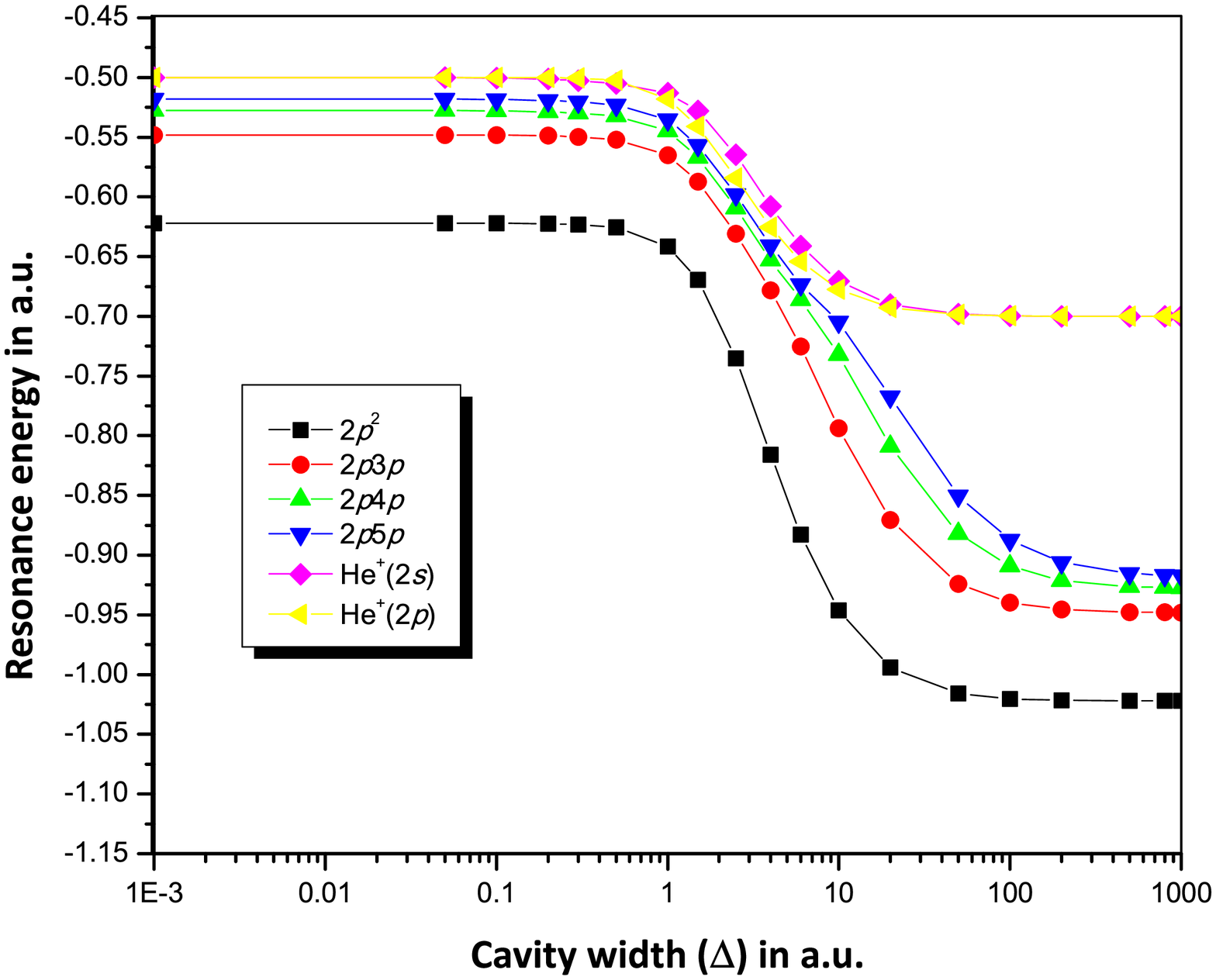}
\vspace{-0.5cm}
\caption{The variation of resonance energies ($E_{r}$) of $2pnp$ [$n=2-5$] ($^{1}S^{e}$) states and corresponding $2s$ and $2p$ threshold energies with the cavity width ($\Delta$).}
\end{figure}
\begin{figure}
\includegraphics[scale=0.5]{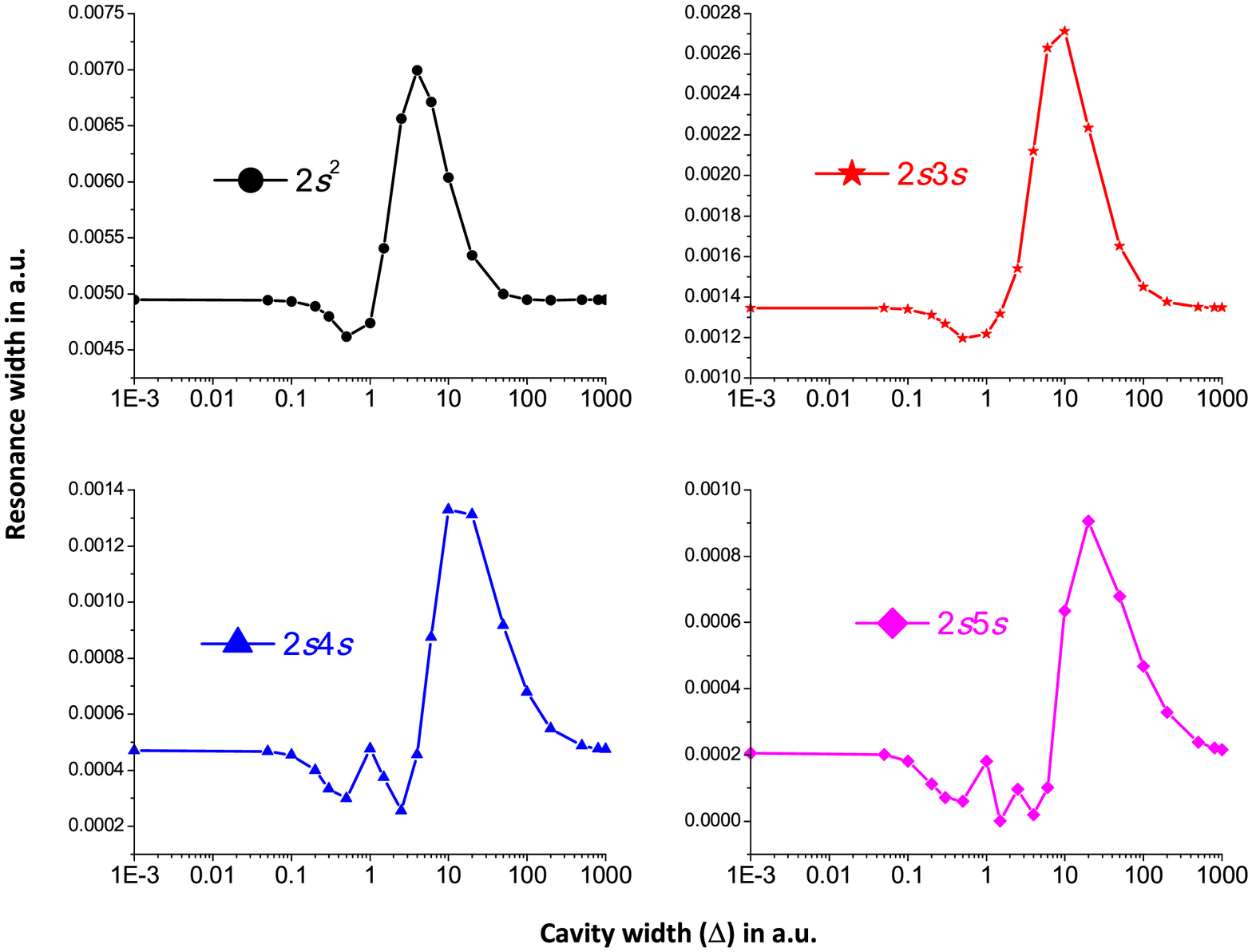}
\vspace{-0.5cm}
\caption{The variation of resonance width ($\Gamma$) of $2sns$ [$n=2-5$] ($^{1}S^{e}$) states with the cavity width ($\Delta$).}
\end{figure}
\begin{figure}
\includegraphics[scale=0.5]{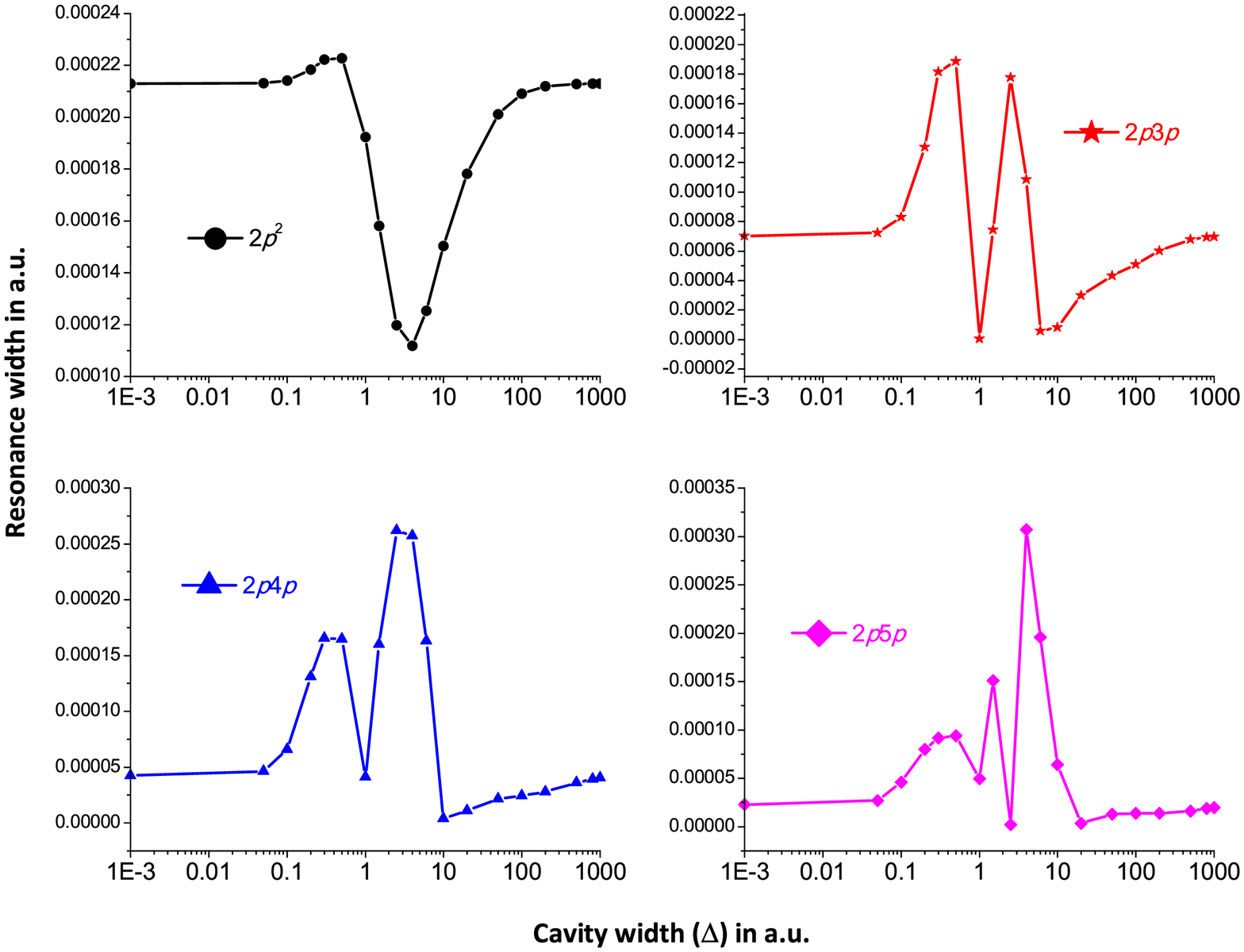}
\vspace{-0.5cm}
\caption{The variation of resonance width ($\Gamma$) of $2pnp$ [$n=2-5$] ($^{1}S^{e}$) states with the cavity width ($\Delta$).}
\end{figure}
\end{document}